\documentclass[letterpaper, 10 pt, conference]{ieeeconf}

\IEEEoverridecommandlockouts                              

\usepackage{graphics} 
\usepackage{epsfig} 
\usepackage{amsmath} 
\usepackage{amssymb}  
\usepackage{flushend}

\title{\LARGE \bf
Distributed averaging for accuracy prediction in networked systems
}

\author{Christel Sirocchi$^{1}$ and Alessandro Bogliolo$^{2}$
\thanks{$^{1}$Christel Sirocchi is a PhD candidate in Research Methods in Science and Technology at the Department of Pure and Applied Sciences, University of Urbino (Italy) {\tt\small c.sirocchi2@campus.uniurb.it}}%
\thanks{$^{2}$Alessandro Bogliolo is a Full Professor of Computer Science at the Department of Pure and Applied Sciences, University of Urbino (Italy) {\tt\small alessandro.bogliolo@uniurb.it}}%
}

\begin{document}

\maketitle
\thispagestyle{empty}
\pagestyle{empty}

\begin{abstract}

Distributed averaging is among the most relevant cooperative control problems, with applications in sensor and robotic networks, distributed signal processing, data fusion, and load balancing. Consensus and gossip algorithms have been investigated and successfully deployed in multi-agent systems to perform distributed averaging in synchronous and asynchronous settings. This study proposes a heuristic approach to estimate the convergence rate of averaging algorithms in a distributed manner,  relying on the computation and propagation of local graph metrics while entailing simple data elaboration and small message passing. The protocol enables nodes to predict the time (or the number of interactions) needed to estimate the global average with the desired accuracy. Consequently, nodes can make informed decisions on their use of measured and estimated data while gaining awareness of the global structure of the network, as well as their role in it. The study presents relevant applications to outliers identification and performance evaluation in switching topologies.

\end{abstract}

\section{INTRODUCTION}
Distributed averaging is an instance of distributed computation aiming to determine the global average of a set of values by iterating local calculations. It has been extensively studied as the primary tool for solving cooperative control problems in multi-agent systems~\cite{martinovic2022cooperative} such as distributed sensor networks, micro-grids, transport networks, power distribution systems, and biological networks~\cite{loizou2021revisiting}. In these settings, the value of an aggregate function over the entire network data is often more relevant than individual data at nodes. Networks of temperature sensors, for instance, are generally deployed to assess the average temperature in a given area. Similarly, peer-to-peer systems are primarily interested in calculating the average size of stored files~\cite{kempe2003gossip}. Other collective behaviours leveraging distributed averaging include formation control of autonomous vehicles, network synchronisation, automated traffic networks, cooperative output regulation, and containment control~\cite{trinh2017theory}.\\
The performance of a distributed averaging protocol is generally evaluated as the resources required to obtain an estimate of the global average with the desired level of accuracy. Performance analysis is fundamental to networked systems, as they typically suffer limitations in terms of communication bandwidth, memory, and computational power~\cite{sirocchi2023community}. Previous work has focused on deriving theoretical performance guarantees in synchronous~\cite{olfati2007consensus} and asynchronous~\cite{boyd2005gossip} models and identified a dependency on the eigenvalues of the matrix characterising the algorithm. However, these results cannot help make local and accurate performance predictions, as they identify wide performance intervals and require knowledge of the entire network structure for eigenvalues calculations. In contrast, this work proposes a strategy to estimate global properties of the graph and the performance of the averaging algorithm in a distributed manner, entailing simple data elaboration and small message passing. As a result, agents can predict the time (or the number of interactions) needed to achieve the desired accuracy, make informed decisions on their use of measured and estimated data, and also gain awareness of the global network structure.\\
The remainder of the paper is organised as follows. Section~\ref{BG} provides some relevant background on averaging algorithms, citing previous results on convergence for synchronous and asynchronous models. The problem is formulated in Section~\ref{PS}, and the proposed strategy is outlined in Section~\ref{MR}. Section~\ref{IE} presents relevant applications, and finally, Section~\ref{CL} provides conclusions and directions for future work.


\section{BACKGROUND}\label{BG}

\subsection{Network Topology}
The communication constraints in networked systems can be conveniently modelled via a graph $G = (V, E)$, where $V$ is the vertex set of $n$ nodes $v_i$, with $i \in I = \{1,\dots, n\}$ and $n \in N$, and $E$ is the edge set $E \subseteq V \times V$ of the pairs $e_{ij} = (v_{i}, v_{j})$, so that there is an edge between nodes $v_i$ and $v_j$ iff $(v_{i},v_{j}) \in E$. In undirected graphs, edges $e_{ij} \in E$  are unordered pairs ($v_i$, $v_j$) of elements of $V$. All nodes that can transmit information to node $v_i$ are said to be its neighbours and are represented by the set $\Omega_i = \{v_{j} : (v_{i}, v_{j}) \in E\}$. 



\subsection{Distributed average}
Let $x_{i}$ denote the value of node $v_i$, representing a physical quantity such as position, temperature, light intensity or voltage, and $\textbf{x} = (x_1, ..., x_n)^{T}$ the vector of values so that the $i^{th}$ component of $\textbf{x}$ is the value at node $v_{i}$. The nodes $v_i$ and $v_j$ are said to agree in a network iff $x_{i} = x_{j}$. All nodes in $G$ are in agreement or have reached a consensus iff $x_{i} = x_{j} \, \forall \, i,j \in I$. This agreement space can also be expressed as $\textbf{x} = \alpha \textbf{1}$ where $\textbf{1} = (1,...,1)^T$ and $\alpha \in R$ is the collective decision value of the nodes. The system is said to reach asymptotic consensus if all nodes asymptotically converge to $\alpha$, i.e.
$$\lim_{t \to +\infty} x_i(t) = \alpha , \forall i \in I.$$
A consensus algorithm (or protocol) is an interaction rule that specifies the information exchange between an agent and its neighbours to reach a consensus~\cite{olfati2007consensus}, i.e. to asymptotically converge to the agreement space~\cite{sundaram2008distributed}. One of the benefits of using linear iteration-based schemes is that each node only transmits a single value to each of its neighbours~\cite{sundaram2008distributed}. In discrete time, the consensus protocol is
\begin{equation}
\label{eqn:iteration}
\textbf{x}(k) = \textbf{W}(k) \, \textbf{x}(k-1)
\end{equation}
where $\textbf{x}(k)$ is the vector of values at the end of time slot $k$, and $\textbf{W}$ is the $n \times n$ matrix of averaging weights. When the consensus value corresponds to the average of all initial values, i.e. $\alpha = \frac{1}{n}\sum_{i = 1}^{n}x_i(0)$, the system is said to perform distributed averaging. In consensus protocols with time-invariant weight matrix $\textbf{W}$, the linear iteration implies \begin{equation}
\label{eqn:multiple}
\textbf{x}(k) = \textbf{W}^k \, \textbf{x}(0).
\end{equation}
To achieve asymptotic average consensus regardless of the initial values $\textbf{x}(0)$
\begin{equation}
\label{eqn:limit}
\lim_{t \to +\infty} \textbf{W}^t = \frac{\textbf{1}\textbf{1}^T}{n}
\end{equation}
which follows from
\begin{equation}
\lim_{t \to +\infty} \textbf{x}(t) = \lim_{t \to +\infty} \textbf{W}^t\textbf{x}(0) = \frac{\textbf{1}\textbf{1}^T}{n} \; \textbf{x}(0).
\end{equation}
Eq.~\ref{eqn:limit} holds iff the following three properties are satisfied:
\begin{equation}
\textbf{1}^T\textbf{W} = \textbf{1}^T
\label{eqn:con1}
\end{equation}
i.e. \textbf{1} is a left eigenvector of $\textbf{W}$ associated with the eigenvalue 1, implying that $ \textbf{1}^T\textbf{x}(k+1) = \textbf{1}^T\textbf{x}(k)$ $\forall \;k$, i.e., the sum, and therefore the average, of the vector of node values is preserved at each step;
\begin{equation}
\textbf{W}\;\textbf{1} = \textbf{1}
\label{eqn:con2}
\end{equation}
i.e. \textbf{1} is also a right eigenvector of $\textbf{W}$ associated with the eigenvalue 1, meaning that \textbf{1}, or any vector with constant entries, is a fixed point for the linear iteration; 
\begin{equation}
\rho(\textbf{W} - \frac{\textbf{1}\textbf{1}^T}{n}) < 1
\label{eqn:con3}
\end{equation}
where $\rho(\cdot)$ denotes the spectral radius of a matrix, which, combined with the first two conditions, states that 1 is a simple eigenvalue of $\textbf{W}$, and all other eigenvalues have a magnitude strictly less than 1~\cite{xiao2004fast}.

The disagreement vector $\boldsymbol{\delta}(k)$ quantifies the distance from consensus at time slot $k$ and can be computed as
\begin{equation}
\boldsymbol{\delta}(k) = \textbf{x}(k) - \alpha \textbf{1}.
\end{equation} 
Notably, this vector evolves according to the same linear system as the vector of values:
\begin{equation}
\boldsymbol{\delta}(k) = \textbf{W}(k)\boldsymbol{\delta}(k-1).
\end{equation} 
Iterative distributed averaging algorithms are often classified based on the adopted time model. Gossip algorithms realise asynchronous averaging schemes so that a single pair of neighbouring nodes interact at each time $k$, setting their values to the average of their previous values~\cite{boyd2006randomized}. In contrast, consensus algorithms implement a synchronous model where time is commonly slotted across nodes, and all nodes simultaneously update their values with a linear combination of the values of their neighbours at discrete times $k$. Gossip protocols are more suited to model real networks and are generally more accessible to implement for the lack of synchronisation requirements, unrealistic for most applications~\cite{oliva2019gossip}, but are harder to characterise mathematically due to the added randomness of the neighbour selection.



\subsection{Gossip algorithms}
In asynchronous gossip protocols, a single node is active at each time slot $k$ and selects one of its neighbours for interaction according to a given criterion. The $n \times n$ probability matrix $\textbf{P} = [p_{ij}]$ prescribes the probability $p_{ij}$ that the node $v_i$ selects node $v_j$, with $p_{ij} = 0$ if $(v_i,v_j) \not\in E$ due to the constraints of only interacting with neighbours. For instance, in random neighbour selection, where all neighbours are equally likely to be chosen, the matrix $\textbf{P}$ is [$p_{ij}$] such that $p_{ij} = 1/|\Omega_i|$ $\forall v_j \in \Omega_i$ and 0 otherwise. A node $v_{i}$ interacts with node $v_{j}$ at time slot $k$ with probability $\frac{p_{ij}}{n}$, which is the joint probability that $v_i$ is active at time slot $k$ ($p = \frac{1}{n}$) and selects node $v_{j}$ for interaction ($p = p_{ij}$). The weight matrix $\textbf{W}_{ij}$ of this averaging scheme has elements

\begin{equation}
w_{kl}(t) =
\begin{cases}
\frac{1}{2} & \text{if } k, l \in \{i,j\}\\
1 & \text{if } k = l, k \not\in \{i,j\}\\
0 & \text{otherwise.}
\end{cases} 
\end{equation} 

This is equivalent to nodes $v_{i}$ and $v_{j}$ setting their values to the average of their current values, leaving the others unchanged. The matrix $\textbf{W}$ generally changes over time, as different pairs interact at each time slot. The averaging process is thus defined by the sequence of averaging matrices $\{\textbf{W}(k)\}_{k}$
and the vector value at time step $k$ can be computed as
\begin{equation}
\textbf{x}(k) = \textbf{W}(k-1)\textbf{W}(k-2)\text{ .. }\textbf{W}(0)\textbf{x}(0) = \boldsymbol{\phi}(k-1)\textbf{x}(0)
\end{equation} 
Recalling that for independent real-valued random matrices the expected matrix of the product is the product of the expected matrices, then
\begin{equation}
\mathbb{E}(\boldsymbol{\phi}(k)) = \prod_{i=0}^{k} \mathbb{E}(\textbf{W}(i)) = \bar{\textbf{W}}^k
\label{eqn:product}
\end{equation} 
where $\bar{\textbf{W}}$ is the expected weight matrix \begin{equation}
\bar{\textbf{W}} = \sum_{i,j} \frac{p_{ij}}{n}\textbf{W}_{ij}
\end{equation}
most commonly written as
\begin{equation}
\label{eqn:expected}
\bar{\textbf{W}} = \textbf{I} - \frac{1}{2n}\textbf{D} + \frac{\textbf{P} + \textbf{P}^T}{2n},
\end{equation}
\cite{boyd2006randomized} where \textbf{I} is the identity matrix and \textbf{D} is the diagonal matrix with entries
$$ \textbf{D}_i = \sum_{j=1}^{n}[p_{ij} + p_{ji}].$$
By definition, $\bar{\textbf{W}}$ is symmetric and doubly stochastic, i.e., all rows and columns sum up to 1. If the underlying graph is connected and non-bipartite, the expected matrix $\bar{\textbf{W}}$ fulfils all three conditions for convergence (Eq.~\ref{eqn:con1},~\ref{eqn:con2},~\ref{eqn:con3}), so the sequence of averaging matrices $\{\textbf{W}(k)\}_{k}$ drawn independently and uniformly and applied to any initial vector $\textbf{x}(0)$, converges to the vector average $\frac{\textbf{1}\textbf{1}^T}{n}\textbf{x}(0)$~\cite{boyd2006randomized}. Notably, the second largest eigenvalue of the matrix $\bar{\textbf{W}}$ determines the performance of the gossip scheme~\cite{boyd2005gossip}. 






\subsection{Consensus algorithms}

Consensus algorithms perform linear iterations where each node updates its value to a weighted average of its own previous value and those of its neighbours in a synchronous manner. The simple consensus scheme to reach an agreement regarding the state of $n$ agents with dynamics $\dot x_i = u_i$ can be expressed as the linear system
\begin{equation}
\dot x_i(k) = \sum_{v_j \in \Omega_i} a_{ij}(x_j(k) - x_i(k))
\end{equation}
where $i \in I$, $k$ is the discrete time index, and $a_{ij}$ is a weight associated to the edge $(v_i,v_j)$~\cite{olfati2004consensus}. Setting $a_{ij} = 0$ for $v_j \notin \Omega_i$, this iteration can be written as Eq.~\ref{eqn:iteration} and~\ref{eqn:multiple}. The collective dynamics of the group of agents can be expressed in compact form as
\begin{equation}
\label{eqn:dynamics}
\boldsymbol{\dot x} = -\boldsymbol{\mathcal{L}} \; \textbf{x},
\end{equation}
so that $\textbf{W} = \textbf{I} - \boldsymbol{\mathcal{L}}$, with identical disagreement dynamics
\begin{equation}
\boldsymbol{\dot \delta} = -\boldsymbol{\mathcal{L}} \; \boldsymbol{\delta}.
\end{equation}
A state in the form $\alpha\textbf{1}$, where all nodes agree, is an asymptotically stable equilibrium of the dynamic system in Eq.~\ref{eqn:dynamics} because 
\begin{equation} 
-\boldsymbol{\mathcal{L}} \; \alpha\textbf{1} = \textbf{0}.
\end{equation}
The consensus algorithm asymptotically converges to the agreement space provided that $\boldsymbol{\mathcal{L}}$ is a positive semidefinite matrix and $\alpha\textbf{1}$ is the only equilibrium of the system. It was shown that, in connected undirected graphs, the equilibrium is unique and corresponds to the vector average. Moreover, the convergence rate of the consensus algorithm depends on the second smallest eigenvalue of the Laplacian matrix defined on $G$~\cite{olfati2007consensus}.\\

From Eq.~\ref{eqn:product}, it follows that the performance of a gossip protocol with expected weight matrix $\bar{\textbf{W}}$, as defined in Eq.~\ref{eqn:expected}, converges in expectation with that of a consensus scheme with time-invariant weight matrix $\bar{\textbf{W}}$. The agents dynamics of such system is
\begin{equation} 
\dot x_i(t) = \sum_{v_j \in \Omega_i} \frac{p_{ij} + p_{ji}}{2n}(x_j(t) - x_i(t))
\end{equation}
while the collective dynamics is given by Eq.~\ref{eqn:dynamics} where
\begin{equation*} 
\boldsymbol{\mathcal{L}} = \frac{\textbf{D}}{2n} - \frac{\textbf{P} + \textbf{P}^T}{2n}.
\end{equation*}
So defined, $\boldsymbol{\mathcal{L}}$ is positive-semidefinite, as it is symmetric and diagonally dominant, and has all row-sums and column-sums equal to zero, so $\boldsymbol{\mathcal{L}}$ always has a zero eigenvalue corresponding to the eigenvector \textbf{1}. In connected graphs, the vector average is a unique equilibrium for the system, and the algorithm asymptotically converges to the agreement space.





\subsection{Convergence rate and Accuracy}

\begin{figure}[thpb]
  \centering
  \includegraphics[scale=0.7]{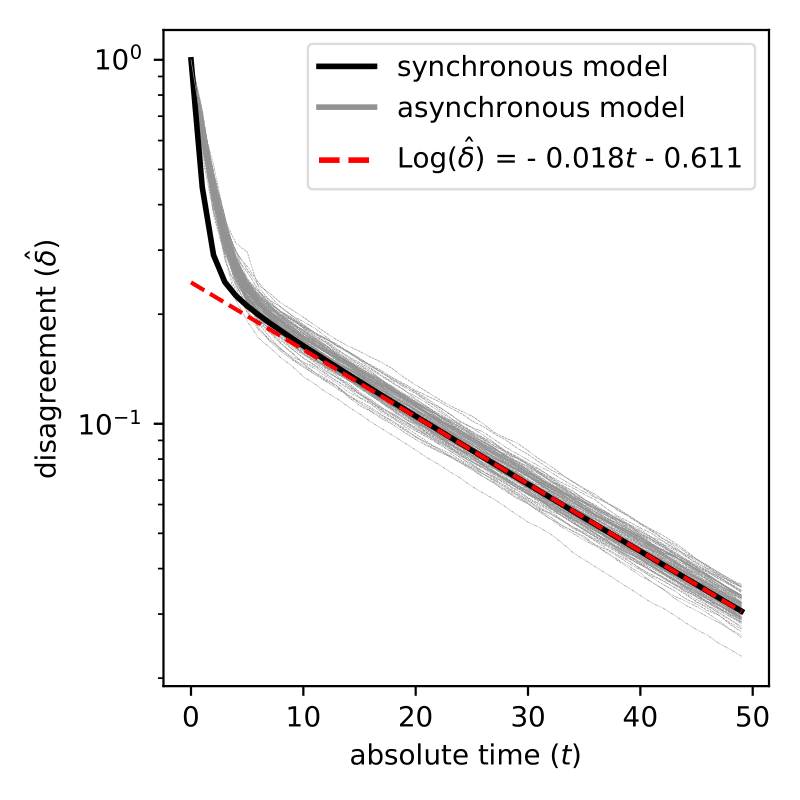}
  \caption{Collective disagreement over time in 100 realisations of an asynchronous averaging protocol with random neighbour selection and the corresponding synchronous scheme, simulated on a geometric random graph of size $n$ = 200 and average degree ${<}\textit{k}{>}$ = 25. The contraction rate $\gamma$ is estimated as 0.018.}
  \label{fig:synch}
\end{figure}

The convergence of the distributed iteration is governed by the product of matrices, each of which satisfies certain communication constraints imposed by the graph topology and the consensus/gossip criterion. Let $\hat{\delta}(t)$ be the collective disagreement of the estimates at time $t$ normalised by the initial values
\begin{equation}
\hat{\delta}(t) =  \frac{\| \boldsymbol{\delta}(t) \|}{\| \textbf{x}(0) \|},
\end{equation}
where $\| . \|$ is the $l_{2}$ norm of the vector. Notably, the $l_{2}$ norm of the disagreement vector $\boldsymbol{\delta}$ corresponds to the standard deviation of the vector entries scaled by the network size
\begin{equation} 
\sigma(t)=  \frac{\| \boldsymbol{\delta}(t) \|}{\sqrt{n}}.
\label{eqn:std}
\end{equation}
Numerical and theoretical results for synchronous and asynchronous averaging schemes indicate that the logarithm of the collective disagreement $\hat{\delta}(t)$ decreases linearly after a faster transient phase, and that the decreasing rate is deterministic and independent of the initial measurements $\textbf{x}(0)$~\cite{denantes2008distributed}. Hence, a contraction rate $\gamma$ can be defined as the angular coefficient of the linear stationary regime and used to characterise the algorithm performance~\cite{sirocchi2022topological}:
\begin{equation} 
Log(\hat{\delta}) = - \gamma \; t.
\end{equation}
The accuracy $R$ of the estimates at time $t$ indicates by how many times the collective disagreement is reduced with respect to the initial disagreement
\begin{equation} 
R(t) = - Log(\frac{\hat{\delta}(t)}{\hat{\delta}(0)}) = - Log(\frac{\| \boldsymbol{\delta}(t) \|}{\| \boldsymbol{\delta}(0) \|})
\label{eqn:ratio}
\end{equation}
so the time taken to achieve the desired level of accuracy can be computed as
\begin{equation} 
t = R/\gamma
\label{eqn:R}
\end{equation}
If nodes initiate, on average, one interaction per unit of time, the time parameter $t$ approximates the number of interactions per node. Figure~\ref{fig:synch} shows the collective disagreement over time for several realisations of an asynchronous averaging algorithm and the corresponding synchronous scheme, as well as the regression line whose angular coefficient is the chosen metric of convergence rate $\gamma$.

\section{PROBLEM SETUP}\label{PS}

Let us consider a network of agents, each with some initial value $x_i(0)$ representing an opinion or a measurement, implementing a synchronous or asynchronous averaging strategy for reaching an agreement on the global average $\alpha$. Each node is endowed by a parameter $r_i$, indicating the accuracy of its estimated average $x_i(t)$ required to confidently use it, elaborate it, communicate it to a remote server or compare it to its initial measurement. An accuracy parameter $r_i$ equal to $R$ indicates that the node can only use its estimate when the collective disagreement in the system is smaller than the initial disagreement divided by $10^R$. For instance, sensor nodes might confidently use their estimates when the group disagreement has been significantly reduced (e.g. $R$ = 4), but nodes with control responsibilities might require estimates of many orders of magnitude more accurate (e.g. $R$ = 8).  

\section{PROPOSED STRATEGY and MAIN RESULTS}\label{MR}
In the considered setting, the global structure of the graph is unknown to the individual nodes and can change over time. However, nodes can exchange information with their immediate neighbours to gain awareness of their surroundings and compute local metrics, namely their degree, clustering coefficient and local efficiency. Simulations deployed on over 12000 graphs show that a linear combination of averages of local metrics highly predicts the convergence rate of distributed averaging algorithms in both synchronous and asynchronous time models. Consequently, it is suggested that nodes compute local metrics and estimate their average across the network by distributed averaging, exploiting the same principle used to calculate the average of measured quantities. Nodes then employ averaged local metrics to estimate the algorithm convergence rate and make predictions of the time (or number of interactions) needed to achieve the desired accuracy so that nodes use their estimate only when confident of their quality. Although a combination of more local metrics or more complex models could provide higher explanatory power and better predictions, the limited memory and computational capabilities available to nodes motivate the choice of a linear model of few parameters.\

Besides enabling performance predictions, local metric averages offer nodes a glimpse into the global properties of the network. The average node degree, for instance, is known to play a universal role in cooperation and robustness~\cite{shames2012distributed}. Furthermore, by comparing its local metrics with the population averages, a node gains awareness of its role in the network without computing costly centrality measures. Nodes having a degree significantly lower than the average degree recognise that they limit the overall performance of the graph and are encouraged, if possible, to create new connections. Highly clustered nodes are known to reduce communication efficiency by propagating redundant information and can be programmed to remove or rewire some of their connections to increase performance~\cite{brust2007small}.

\subsection{Local Graph Metrics}
The degree of a node $v_i$, here denoted by $k(v_{i})$, is the cardinality of the neighbour set $\Omega_i$ and quantifies its connections within the network. The clustering coefficient $cl(v_{i})$ is defined as the number of triangles passing through the node $T(v_i)$ divided by the number of possible triangles
\begin{equation*} 
cl(v_i) = \frac{2T(v_i)}{k(v_i)(k(v_i)-1)},
\end{equation*}
and is a measure of the degree to which nodes tend to cluster together. The local efficiency \textit{eff}$(v_{i})$ is the average efficiency of all node pairs in the sub-graph induced by the neighbours of $v_i$, where the efficiency of a node pair is the multiplicative inverse of the shortest path distance. It quantifies the resistance to failure on a small scale as it measures how effectively information is exchanged after removing the node~\cite{latora2001efficient}. 
The computation complexity of local metrics largely depends on network density. For very sparse graphs, local metrics can be computed by each node in constant time and in a fully parallel fashion. 


\subsection{Regression Model}

The study generated over 12000 sparse fully-connected undirected networks belonging to different graph families ($\approx$ 1600 Erd\H{o}s-Rényi graphs~\cite{erdHos1960evolution}, 1600 geometric random graphs~\cite{penrose2003random}, 4400 small world graphs~\cite{watts1998collective}, 4400 scale-free graphs~\cite{barabasi1999emergence} with adjusted clustering~\cite{holme2002growing}) having sizes ranging from 100 to 1000 nodes and average degree up to 60. Averages of local metrics were calculated for all graphs, and the convergence rates of the asynchronous protocol were estimated through simulations. The adopted gossip scheme realises a random neighbour selection and activates each node at the times of a rate 1 Poisson process so that $n$ interactions take place on average for each unit of time. A regression model in the form
\begin{equation} 
 \gamma = a\,{<}\textit{k}{>} + b\,{<}\textit{cl}{>} + c{<}\,\textit{eff}{>} + d,
\end{equation}
where ${<}{>}$ indicates the average of the corresponding local metric, records an r-squared of 0.92, confirming that local metrics alone predict the algorithm convergence rate with high accuracy. The negative coefficient $b$ suggests that highly clustered areas of the graph are less effective in propagating estimates because nodes are more likely to have shared neighbours passing redundant information. The positive coefficients $a$ and $c$ confirm the intuitive notions that more connections and more efficient information flows promote convergence.\

The proposed approach can be implemented with any probability matrix $\textbf{P}$ as long as the corresponding expected weight matrix satisfies the abovementioned convergence criteria. As seen in Section~\ref{BG}, results for the asynchronous averaging scheme extend to the synchronous counterpart as the two algorithms converge in expectation for certain choices of $\textbf{W}$. A faster initial phase characterises the consensus protocol, as seen in Figure~\ref{fig:synch}, so regression models based on asynchronous simulations might offer conservative estimates for the synchronous case.

\begin{figure}[thpb]
  \centering
  \includegraphics[scale=0.5]{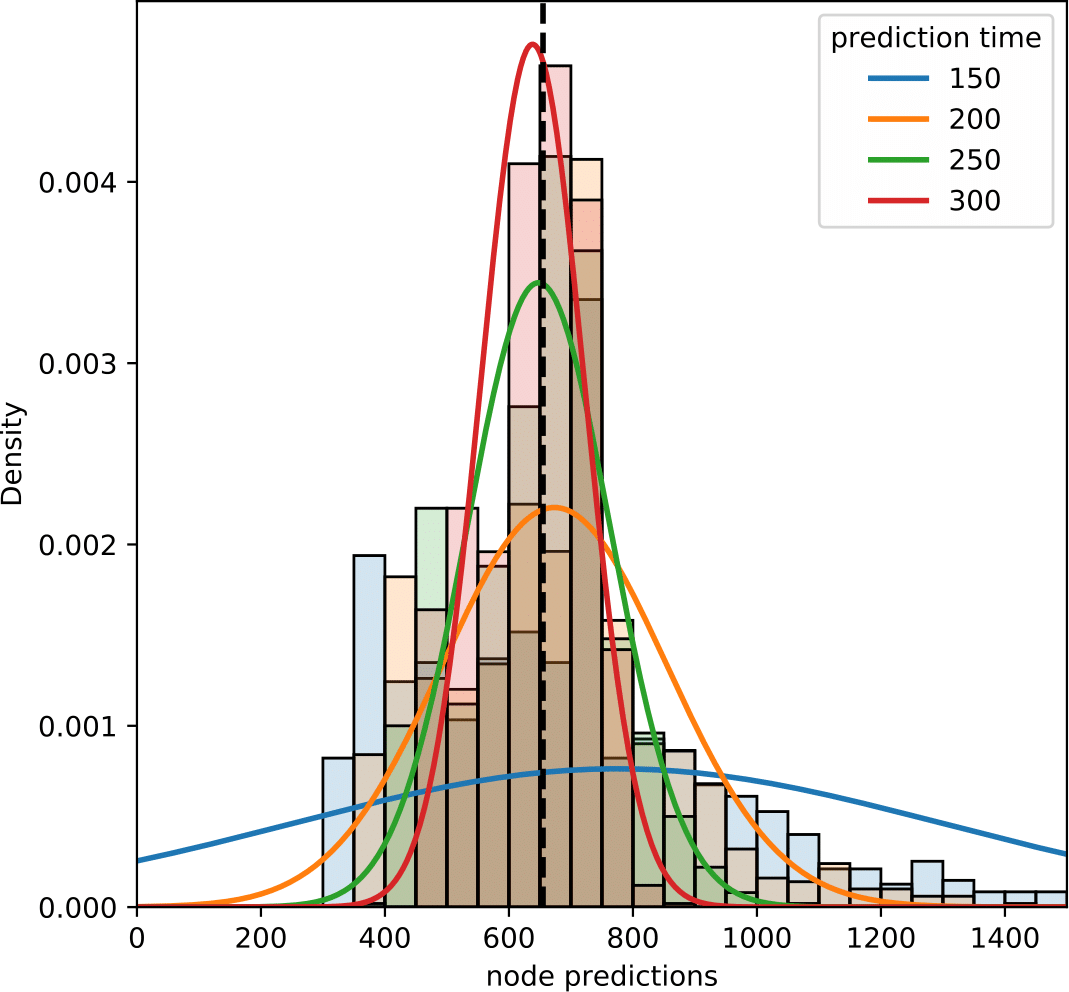}
  \caption{Distribution of node predictions of time needed to achieve accuracy $R$ = 3  in a geometric random graph (size $n$ = 1000, radius $r$ = 0.07) at different time points (150, 200, 250, 300), and the corresponding best-fit normal distributions. The vertical black line indicates the average time at which the desired level of accuracy was achieved in 100 runs of the asynchronous averaging protocol ($t$ = 650). Predictions are accurate and timely as 95\% of them are found in the 650$\pm$50 interval at time $t$ = 300.}
  \label{fig:GRR}
\end{figure}

\subsection{Prediction accuracy}

Node predictions of the time needed to achieve the desired accuracy $R$ are reliable and improve with time, as they normally distribute around a value very close to the actual time, with variance decreasing over time. Necessarily, the accuracy of predictions, like that of the estimates, depends on the algorithm performance, so faster convergence results in more precise time predictions. However, the numerical experiments found that accurate predictions come in a timely manner, as they stabilise to a value of $t$ well before that time elapses, even in the least performing graphs. In Figure~\ref{fig:GRR}, already at $t = 300$, most predictions fall in a narrow interval around the actual value $t = 650$.


\begin{figure}[thpb]
  \centering
  \includegraphics[scale=0.58]{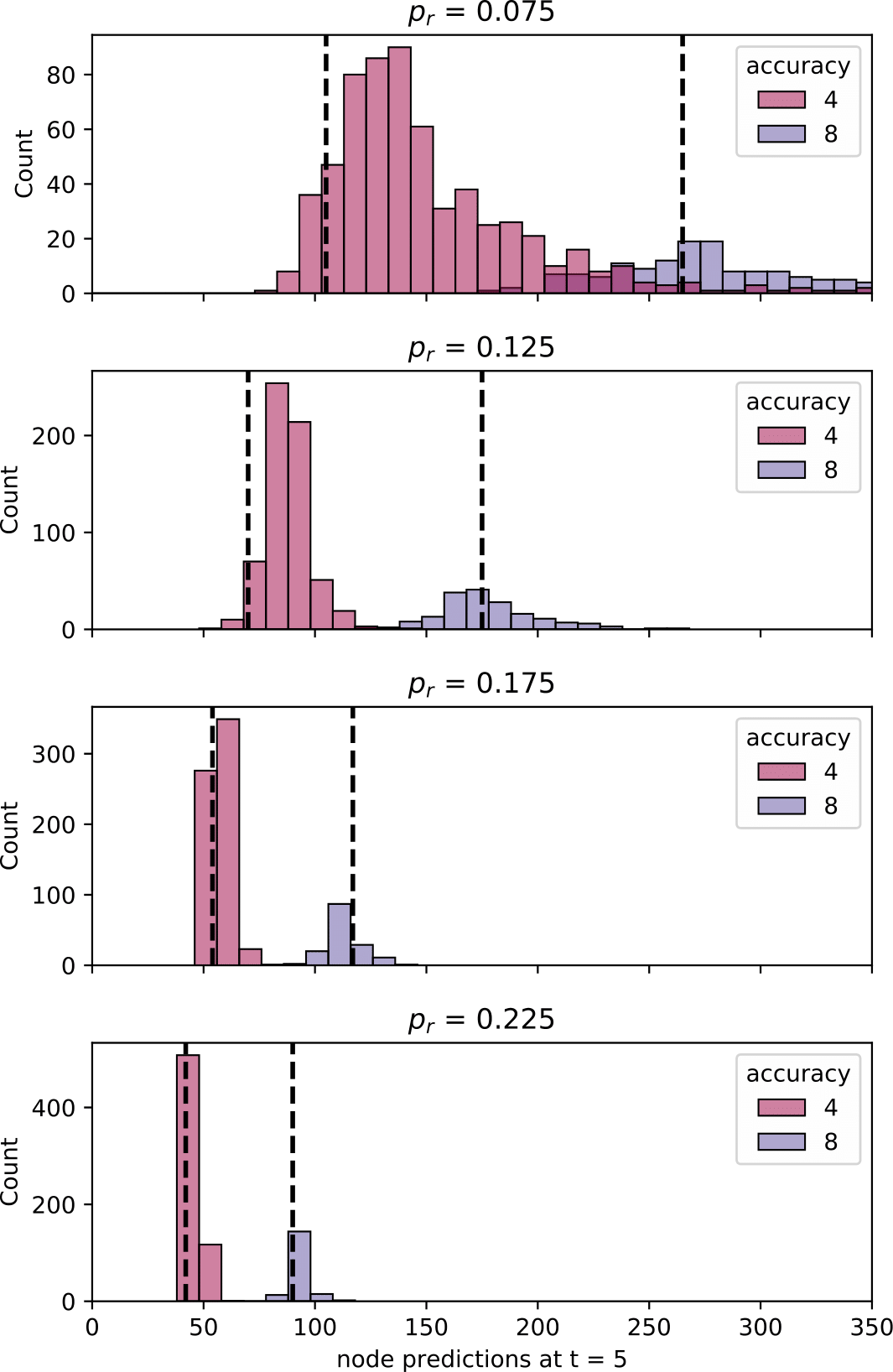}
  \caption{Distribution of node predictions in small world graphs, generated according to the Watts-Strogatz model with size $n$ = 800, average degree ${<}\textit{k}{>}$ = 16, and increasing rewiring probability $p_r$. For all graphs, 80\% of nodes have $R$ = 4, while the remaining 20\% $R$ = 8, modelling a network of agents with different accuracy requirements. The vertical lines represent the average times at which the desired level of accuracy was achieved in 100 realisations of the asynchronous averaging protocol.}
  \label{fig:SW}
\end{figure}

\begin{figure}[thpb]
  \centering
  \includegraphics[scale=0.6]{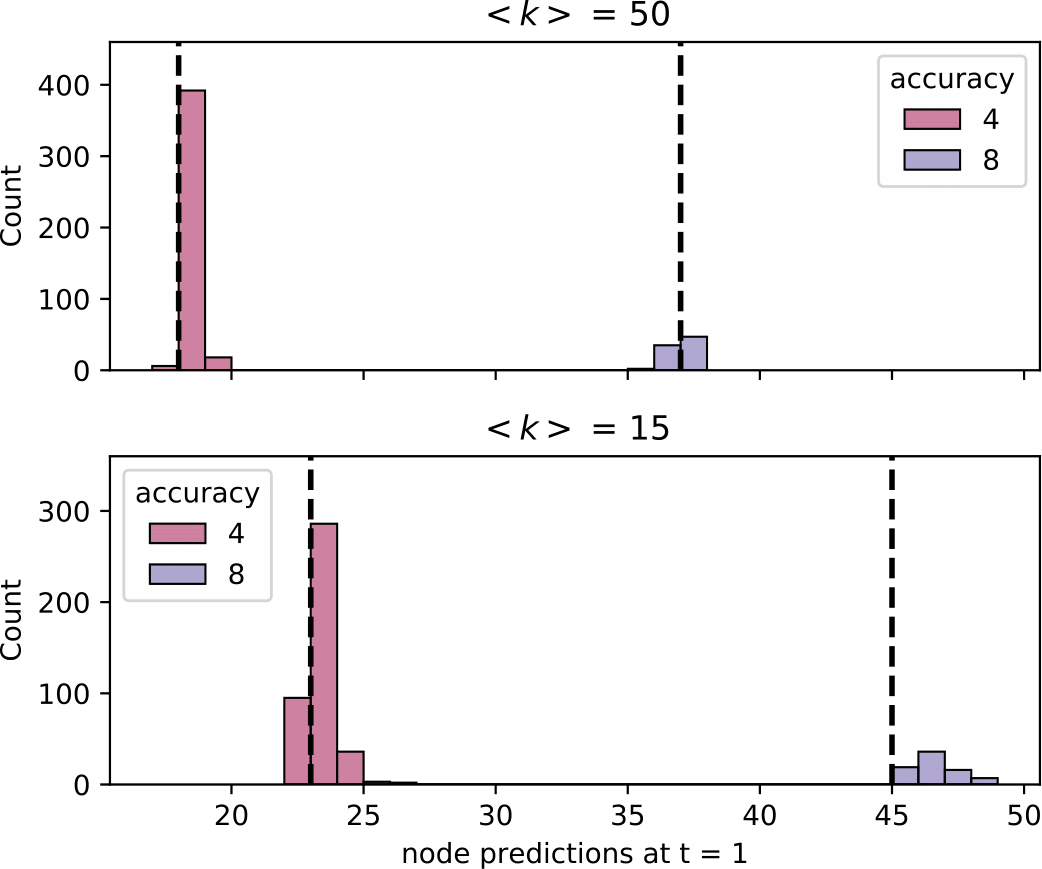}
  \caption{Distribution of node predictions in Erd\H{o}s-Rényi graphs having fixed size $n$ = 500 and decreasing average degree ${<}\textit{k}{>}$ due to link failure (from 50 to 15). The vertical lines represent the average times at which the desired level of accuracy was achieved in 100 realisations of the asynchronous averaging protocol.}
  \label{fig:ER}
\end{figure}

\section{ILLUSTRATIVE EXAMPLES}\label{IE}
\subsection{Topology changes}
Distributed averaging algorithms have been widely investigated in static topologies, characterised by fixed and reliable communication links throughout the observed time. This assumption is generally unrealistic for real networks, where the topology can differ in subsequent executions of the algorithm due to communication interference or agents changing positions~\cite{patterson2010convergence}. However, agents are generally notified solely of changes in their immediate neighbourhood and are unable to assess topology changes on a large scale. The propagation and averaging of local metrics constitute a valuable tool for nodes to evaluate convergence rate in the current communication network, especially in the event of rewiring and link failure, which significantly affects performance. 

\subsubsection{Rewiring}

The convergence rate of averaging protocols can be dramatically increased without adding new links or nodes by means of \textit{random rewiring}~\cite{watts1998collective}, leading to the design of small-world networks for ultra-fast consensus~\cite{olfati2005ultrafast}. The proposed approach enables nodes to estimate the convergence rate of a network undergoing random rewiring at any given time. Figure~\ref{fig:SW} shows subsequent time predictions computed on a small world graph with increasing probability of random rewiring.

\subsubsection{Link failure}

Erd\H{o}s-Rényi graphs are generally adopted to model scenarios where edges fail with equal probability $f$~\cite{hatano2005agreement}. If the failure probability varies over time due to changes in the network communication medium, the graph topology is defined by a series of Erd\H{o}s-Rényi graphs, each with edge probability $1-f$. The proposed method offers nodes a tool to compute the convergence rate and, thus, assess the communication efficiency at any given time. Figure~\ref{fig:ER} shows how predictions capture the lower convergence rate of a network that has lost over 70\% of its links.  

\subsection{Anomaly detection}
Distributed averaging in networked systems allows individual nodes to evaluate how their sensing environment differs from that of the other nodes by comparing their measured value with their estimate of the group average. An anomalous value or outlier for the population can be defined as any measured quantity that is distant from the global average of a given amount $M$ or a certain number of standard deviations $m$. The corresponding node can then raise an alarm to inform a remote control centre of the unexpected measurement, perform a corrective action, limit its interactions, or signal the neighbouring nodes of the lower reliability of its value. 

\subsubsection{Alarm system}
\begin{figure*}[ht]
  \centering
  \includegraphics[width=0.85\textwidth]{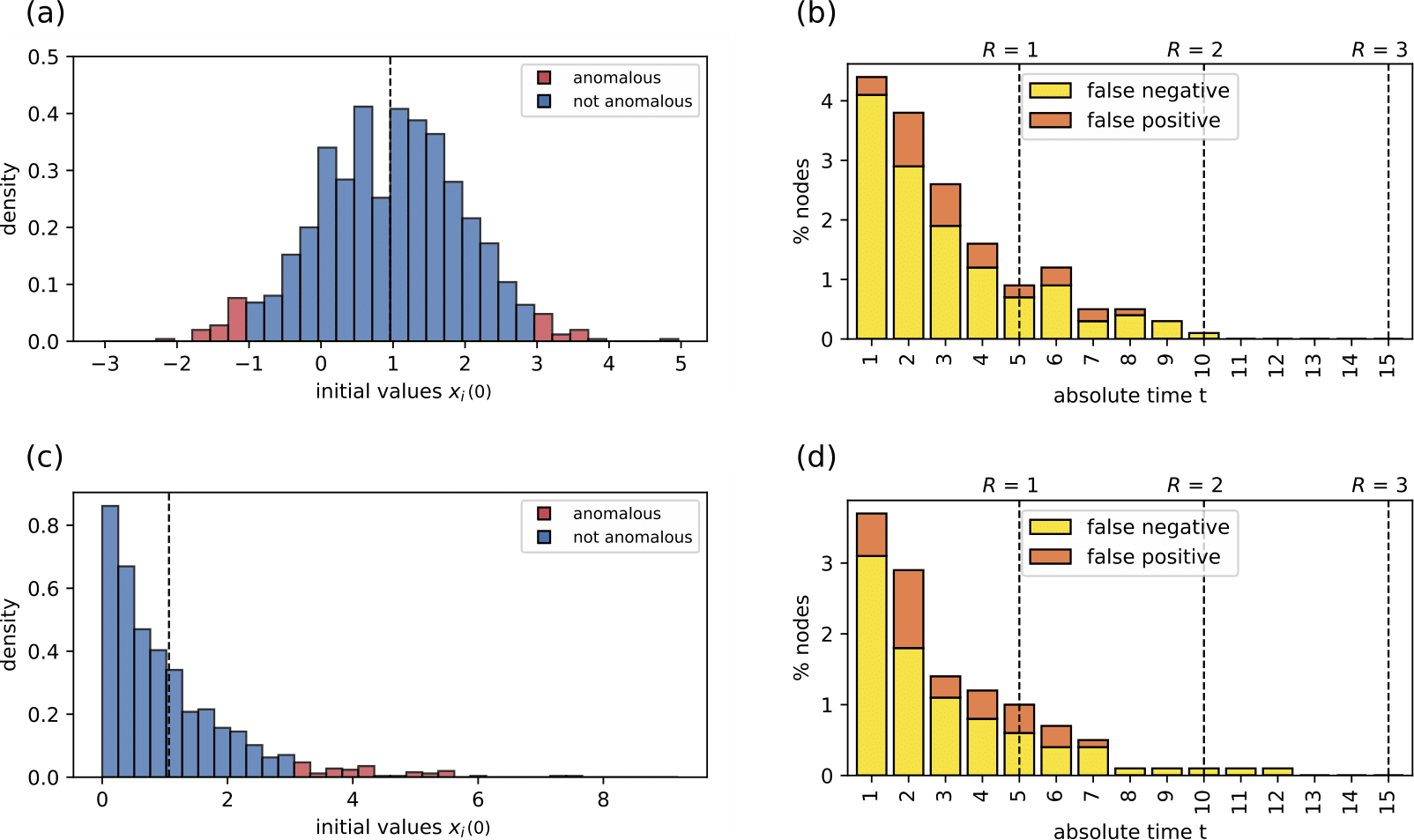}
  \caption{(a) Distribution of initial values $x_i(0)$ sampled from a normal distribution ($\mu = 1$, $\sigma = 1$) and considered anomalous if $|\alpha - x_i(0)| > M$, with $M = 2$. (b) Percentage of nodes making a classification error (false positive or false negative) at time $t$ when comparing their measured value $x_i(0)$ with their average estimate $x_i(t)$. The vertical dotted lines indicate the time at which levels of accuracy $R = 1, 2, 3$ are attained. All nodes evaluate their measurement correctly at $R = 3$. (c) Distribution of initial values $x_i(0)$ sampled from a gamma distribution (shape $k = 1$, scale $\theta = 1$) with $M = 2$, (d) and corresponding classification errors over time. The communication network is an Erd\H{o}s-Rényi graph ($n= 1000, {<}\textit{k}{>} = 10$), and the convergence rate of the averaging gossip algorithm is $ \gamma \approx$ 1/5, so that the collective disagreement is reduced by a factor 10 every 5 units of time.}
  \label{fig:ER2}
\end{figure*}

In networks of sensing devices, any measurement distant at least $M$ from the global average $\alpha$ can be considered anomalous by the system and trigger an alarm. Each node can deploy the proposed approach to evaluate the time $t$ needed to attain its desired accuracy $r_i$ and only then compare its measured value $x_i(0)$ with its estimated average $x_i(t)$. If $|x_i(0) - x_i(t)|> M$, the measurement is labelled as anomalous, and the node activates a response. This procedure is subject to \textit{false negatives}, which do not detect anomalous measurements, and \textit{false positives}, where regular values are erroneously detected as anomalous, because of the differences between the actual global average $\alpha$ and its local estimate at time $t$, $x_i(t)$. Notably, nodes face a trade-off between timely detection and classification error, meaning that a lower value of $r_i$ enables faster but less accurate feedback. Figure~\ref{fig:ER2} exemplifies how classification errors decrease as accuracy requirements increase for two initial distributions of $x_i(0)$.

\subsubsection{Outlier identification}
\begin{figure*}[ht]
  \centering
  \includegraphics[width=0.85\textwidth]{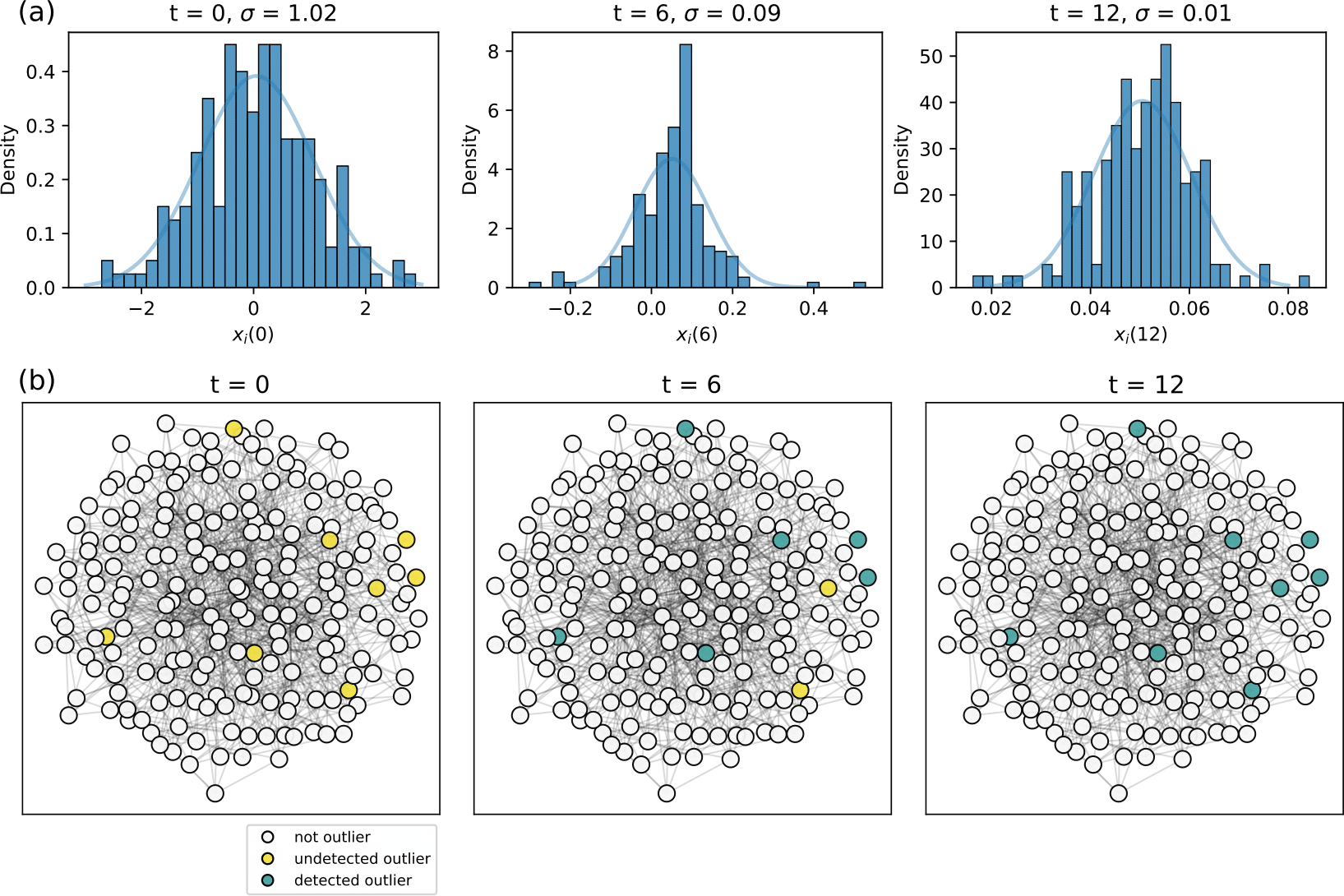}
  \caption{(a) Distribution of the estimates $x_i$ and (b) visual representation of outlier detection at times $t$ = 0, 6, 12. The communication network is a scale-free graph ($n = 200, {<}\textit{k}{>} = 10, cl = 0.01$) with nodes assigned initial values $x_i(0)$ from a normal distribution ($\mu = 0, \sigma = 1$). The convergence rate of the averaging gossip algorithm is $\gamma \approx 1/6$, so the standard deviation of the estimates is expected to become 10 times smaller every 6 units of time. In this scenario, all outliers are detected by $t$ = 12.}
  \label{fig:SF}
\end{figure*}

In sensing systems where measurements are expected to distribute normally with standard deviation $\sigma$, an outlier can be defined as any value distant more than $m$ standard deviations from the group average $\alpha$. From Eq.~\ref{eqn:std},~\ref{eqn:ratio} and~\ref{eqn:R}, it follows that 

\begin{equation}
- Log(\frac{\sigma(t)}{\sigma(0)}) = t \gamma   
\label{eqn:std_est}
\end{equation}

so that, at any time $t$, the standard derivation of the estimates $\sigma(t)$ can be derived from the initial deviation $\sigma(0)$ and the convergence rate of the graph $\gamma$, which can be estimated using the proposed methods. Each node $v_i$ can then keep track of a confidence interval where the group average $\alpha$ is likely to be found. Since estimates are normally distributed, $v_i$ has a high probability of finding the group average within three standard deviations from its estimate, i.e. $\mathbb{P}(\alpha \in ci_i) \approx 0.997$, where $ci_i = [x_i(t) - 3\sigma(t), x_i(t) + 3\sigma(t)]$. If the node initial value $x_i(0)$ is distant at least $m$ standard deviations from this confidence interval, formally
$$\min_{\forall \, c \, \in \, ci_i} | x_i(0) - c | > m\;\sigma(0),$$
the measurement is detected as an outlier for the group, and the corresponding node initiate a response. Figure~\ref{fig:SF} shows that the standard deviation of the estimates evolves according to Eq.~\ref{eqn:std_est}, allowing nodes to evaluate whether their initial value is an outlier for the population.

\section{CONCLUSIONS}\label{CL}


The study proposes a distributed approach to estimate the convergence rate of an averaging scheme deployed on a network. The key idea is to approximate the graph convergence rate with a linear combination of averages of local metrics. Nodes can then be programmed to estimate these parameters by distributed averaging and to implement the regression model in order to compute the convergence rate and the time needed to achieve the desired level of accuracy. The approach enables nodes to make informed decisions on their use of measured and estimated data and gain awareness of the global structure of the network as well as their role in it. Future efforts will be directed toward identifying models able to provide more accurate predictions without increasing memory requirements or communication costs.

\addtolength{\textheight}{-0cm}   



\bibliographystyle{IEEEtran}
\bibliography{root}

\end{document}